\documentstyle[psfig]{kapproc} 

\kluwerbib

\startauthorindex

\begin{document}

\articletitle{Observational Clues to Brown Dwarf Origins}


\author{Ray Jayawardhana}
\affil{Department of Astronomy, University of Michigan, Ann Arbor, MI 48109, U.S
.A.}
\email{rayjay@umich.edu}

\author{Subhanjoy Mohanty}
\affil{Harvard-Smithsonian Center for Astrophysics, Cambridge, MA 02138, U.S.A.}
 

\author{Gibor Basri}
\affil{Department of Astronomy, University of California, Berkeley, CA 94720, U.
S.A.}

\author{David R. Ardila}
\affil{Bloomberg Center, Johns Hopkins University, 
Baltimore, MD 21218, U.S.A.}

\author{Beate Stelzer}
\affil{Osservatorio Astronomico di Palermo, Piazza del Parlamento 1,  
I-90134 Palermo, Italy}

\author{Karl E. Haisch, Jr.}
\affil{Department of Astronomy, 
University of Michigan, Ann Arbor, MI 48109, U.S.A.}

\chaptitlerunninghead{Brown Dwarf Disks and Accretion}

\anxx{Author1\, and Author2}

\begin{abstract}
Over the past year, we have conducted a multi-faceted program to 
investigate the origin and early evolution of brown dwarfs. Using 
high-resolution Keck optical spectra of $\sim$30 objects near and 
below the sub-stellar boundary in several star-forming regions, 
we present compelling evidence for a T Tauri-like accretion phase 
in young brown dwarfs. Our systematic study of infrared 
$L^\prime$-band (3.8$\mu$m) disk excess in $\sim$50 spectroscopically 
confirmed young very low mass objects reveal that a significant 
fraction of brown dwarfs harbor disks at a very young age. Their 
inner disk lifetimes do not appear to be vastly different from 
those of disks around T Tauri stars. Taken together, our findings
are consistent with a common origin for most low-mass stars, 
brown dwarfs and isolated planetary mass objects.
\end{abstract}

\section*{Introduction}
Brown dwarfs, which straddle the mass range between stars and planets, 
appear to be common both in the field and in star-forming regions. Their
ubiquity makes the question of their origin an important one, both for 
our understanding of brown dwarfs themselves as well as for theories on 
the formation of stars and planets.

In the standard framework, a low-mass star forms out of a collapsing 
cloud fragment, and goes through a ``T Tauri phase'', during which it 
accretes material from a surrounding disk, before arriving on the main 
sequence. There is ample observational evidence now to support many key
aspects of this picture for young solar-mass stars. Whether the same 
scenario holds for objects at and below the sub-stellar limit is an 
open question. Studies of {\it young} sub-stellar objects could provide 
valuable clues to address that question. 

To that end, over the past year, we have undertaken a multi-faceted 
study of very low mass (VLM) objects in star-forming regions and their 
immediate circumstellar environment. Here we report on our investigations
of accretion signatures and disk excess in young brown dwarfs. 

\begin{figure}[hb]
\centerline{\psfig{file=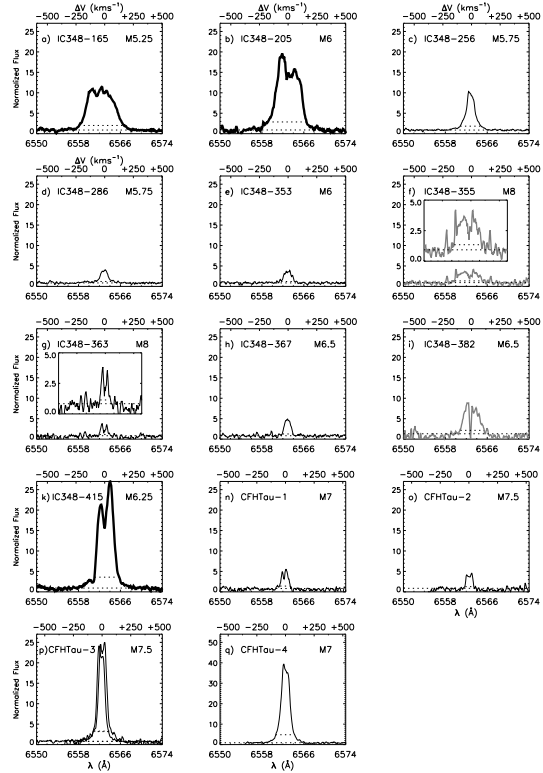,height=4in}}
\caption{H$\alpha$ line profiles of IC 348 and Taurus targets. Spectra 
shown have been smoothed by a 3-pixel boxcar; continuum and full width 
at 10\% of the peak levels are marked by dotted lines.  Thick black lines 
indicate accretors with broad H$\alpha$ as well as CaII and OI emission; 
grey indicates probable accretors, based on the H$\alpha$ profile-shape 
and 10\% full-width.  Insets zoom in on objects with low peak-flux and 
noisy continuua, to clearly show the H$\alpha$ detection.  For CFHT-3, 
two spectra are shown, separated by a year; note the similarly strong 
emission both times.  For CFHT-4, note the change in Y-axis scale; the 
peak flux in this object is much higher than in any other target 
in our sample.}
\end{figure}

\begin{figure}[ht]
\centerline{\psfig{file=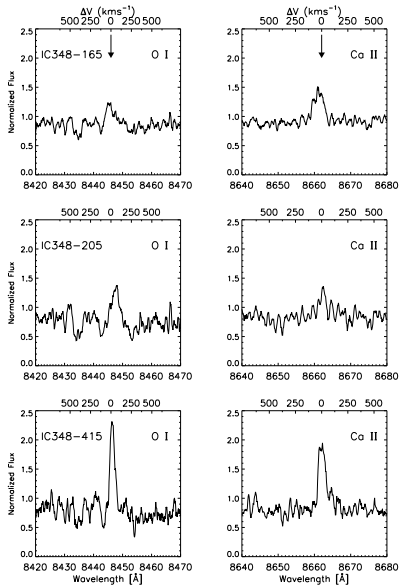,height=3in}}
\caption{Line profiles of OI (8446\AA) and CaII (8662\AA) in three IC 348 
objects.  Arrows indicate line-center.  These spectra have been smoothed 
by a 4-pixel boxcar.}
\end{figure}

\begin{figure}[ht]
\centerline{\psfig{file=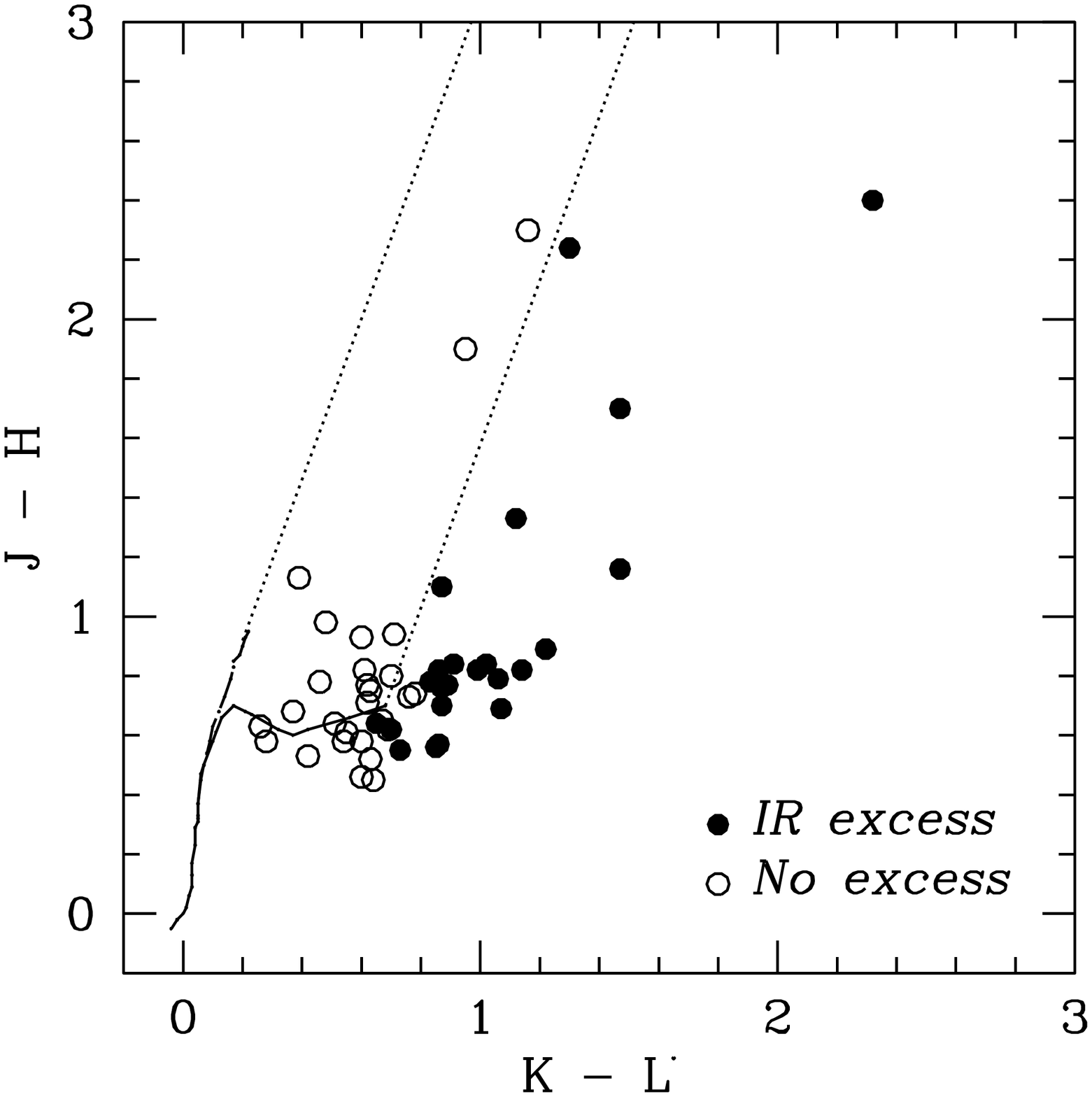,height=3in}}
\caption{$J-H$/$K-L^\prime$ color-color diagram for our target sample. 
Also plotted are the empirical loci of colors for giants (solid) and for 
main-sequence dwarfs (dashed) from Bessell \& Brett (1988) and Leggett 
et al. (2002) and the reddening vectors (dotted). The filled circles 
are stars with  $E(K-L^\prime) >$ 0.2.}
\end{figure}

\section{Accretion Signatures}
The shape and width of the H$\alpha$ emission profile is commonly used to 
discriminate between accretors and non-accretors among T Tauri stars (TTS). 
Stars exhibiting broad, asymmetric H$\alpha$ lines with equivalent width 
larger than 10 \AA~ are generally categorized as classical TTS (CTTS), 
although this threshold value varies with spectral type. Recently White 
\& Basri (2003) have suggested that a full-width $>$ 270 kms$^{-1}$ at 
10\% of the peak emission is a better empirical indicator of accretion, 
independent of spectral type. However, we find that in VLM 
accretors, the H$\alpha$ profile may be somewhat narrower than that in 
higher mass stars. We propose that low accretion rates combined with 
small infall velocities at very low masses can conspire to produce this 
effect, and adopt $\sim$ 200 kms$^{-1}$ as a more appropriate, yet
conservative, threshold (see Jayawardhana, Mohanty \& Basri 2003 for
further discussion). 

We obtained high-resolution Keck optical spectra of $\sim$30 objects spanning
the range of M5--M8 in IC 348, Taurus, Upper Scorpius and $\rho$ Ophiuchus 
star-forming regions. Putting together now all the young objects near or 
below the substellar boundary ($\sim$ M5 and later) with published 
high-resolution optical spectra (Jayawardhana, Mohanty \& Basri 2002; 2003
and references therein), we have 4 objects in $\rho$ Ophiuchus, 
10 in IC 348, 14 in Taurus and 11 in Upper Scorpius. Of these, optical 
spectral signatures of accretion are found in 1 object in $\rho$ Oph (GY 5), 
5 in IC 348 (IC 348-165, 205, 355, 415, 382; Fig. 1), 3 in Taurus 
(CIDA-1, GM Tau, V410 Anon 13) and 
1 in Upper Sco (USco 75), adopting an accretion cutoff of $\sim$ 200 
kms$^{-1}$ in 10\% width of the H$\alpha$ line. The vast majority of 
$\rho$ Oph VLM objects were inaccessible to our optical spectroscopy 
because of significant extinction, 
presumably due to circumstellar as well as interstellar material.  Thus, 
our (small) $\rho$ Oph sample is heavily biased against possible accretors, 
and should not be used to estimate the accreting fraction.  Considering the 
other three clusters, which are much less affected by this bias, we find 
that $\sim$50\% of the VLM objects show disk accretion at an age $\le$ 
2 Myr (IC 348), $\sim$ 20\% at age $\sim$ 3 Myr (Taurus), and $\le$10\% 
by $\sim$ 5 Myr (Upper Sco).  While there are uncertainties in the cluster 
ages, IC 348 is likely to be younger than Taurus and Upper Sco. Thus, we 
appear to be seeing a decrease in the fraction of accreting young sub-stellar 
objects with increasing age.  

Interestingly, three of our $\sim$ M6 IC 348 targets with broad H$\alpha$ 
also harbor broad OI (8446\AA) and CaII (8662\AA) emission (Fig. 2), and 
one shows broad HeI (6678\AA) emission; these features are usually seen 
in strongly accreting classical T Tauri stars. 

Our results constitute the most compelling evidence to date that 
young brown dwarfs undergo a T Tauri-like accretion phase similar to that 
in low-mass stars. 

\section{Disk Excess}
Excesses at infrared wavelengths provide readily a measurable signature 
of dusty disks around late-type objects. Using the ESO Very Large 
Telescope, Keck I and the NASA Infrared Telescope Facility, we have 
carried out a systematic 
study of infrared $L^\prime$-band (3.8$\mu$m) disk excess in a large 
sample of {\it spectroscopically confirmed} objects near and below the 
sub-stellar boundary in several nearby star-forming regions 
(Jayawardhana et al. 2003). Our 
longer-wavelength observations are much better at detecting disk excess 
above the photospheric emission and are less susceptible to the 
effects of disk geometry and extinction corrections than $JHK$ studies
(also see Liu, Najita \& Tokunaga 2003). 

We find disk fractions of 40\%--60\% in IC 348, Chamaeleon I, Taurus and 
Upper Scorpius regions. Based on ISO observations, Natta \& Testi (2001) 
have already shown that Cha H$\alpha$ 1, ChaH$\alpha$ 2 and ChaH$\alpha$ 9 
harbor mid-infrared spectral energy distributions consistent with the presence
of dusty disks. ChaH$\alpha$ 2, which shows a large $K-L^\prime$ excess 
(0.97 mag) in our data is a probable close ($\sim$0.2'') 
binary with roughly equal-mass companions (Neuh\"auser et al. 2002). 
It is possible that a few of our targets harbor infrared companions that
contribute to the measured excess, but this is unlikely in most cases. 
The disk fractions we report for IC 348 and Taurus are lower than those 
found by Liu, Najita \& Tokunaga  (2003). This is primarily because we use 
a more conservative criterion of $K-L^\prime >$ 0.2 for the presence of 
optically thick disks whereas Liu et al. consider all objects with 
$K-L^\prime >$ 0 as harboring disks. In IC 348, our disk fraction is 
comparable to that derived from H$\alpha$ accretion signatures in 
high-resolution optical spectra (Jayawardhana, Mohanty \& Basri 2003).  
However, in Taurus and Upper Sco, which may be slightly older 
at $\sim$2-5 Myrs, we find $K-L^\prime$ excess in $\sim$50\% of the targets 
whereas only three out of 14 Taurus VLM objects and one out of 11 Upper Sco
sources exhibit accretion-like H$\alpha$ (Jayawardhana, Mohanty \& Basri 
2002; 2003). This latter result suggests that dust disks may persist after 
accretion has ceased or been reduced to a trickle, as also suggested by 
Haisch, Lada \& Lada (2001). 

In the somewhat older ($\sim$5 Myr) $\sigma$ Orionis cluster, only about 
a third of the targets show infrared excess. Neither of the two brown 
dwarf candidate members of the $\sim$10-Myr-old TW Hydrae association 
(Gizis 2002) shows excess. Gizis (2002) reported strong H$\alpha$ emission 
(equivalent width $\approx$ 300 \AA) from one of the TW Hydrae objects, 
the M8 dwarf 2MASSW J1207334-393254, and suggested it could be due to 
either accretion or chromospheric activity. Given the 
lack of measurable $K-L^\prime$ excess in this object, accretion 
now appears less likely as the cause of its strong H$\alpha$ emission. 
Our findings in the $\sigma$ Ori and TW Hya associations, albeit for a small 
sample of objects, could mean that the inner disks are clearing out 
by the age of these groups. Similar results have been found for 
T Tauri stars in the TW Hydrae association (Jayawardhana et al. 1999).  

Our results, and those of Muench et al. (2001), Natta et al. (2002), and 
Liu, Najita \& Tokunaga (2003) show that a large fraction of very young brown 
dwarfs harbor near- and mid-infrared excesses consistent with dusty disks.
While the samples are still relatively small, the timescales for inner disk 
depletion do not appear to be vastly different between brown dwarfs and T 
Tauri stars. Far-infrared observations with the {\it Space InfraRed 
Telescope Facility} and/or the {\it Stratospheric Observatory For Infrared 
Astronomy} will be crucial for deriving the sizes of circum-sub-stellar 
disks and providing a more definitive test of the ejection hypothesis 
for the origin of brown dwarfs.

\begin{acknowledgments}
We would like to acknowledge the great cultural significance of Mauna Kea 
for native Hawaiians, and express our gratitude for permission to observe 
from its summit. We thank Geoff Marcy for useful discussions, constant 
encouragement and access to Keck (for the $L^\prime$ observations) and 
Kevin Luhman for valuable assistance during the December 2002 Keck run. 
We are grateful to the staff members 
of the VLT, Keck and IRTF observatories for their outstanding support.
We also thank Fernando Comer\'on and staff at the ESO User Support 
Group for their prompt responses to our queries. This work was 
supported in part by NSF grants AST-0205130 to R.J. and AST-0098468 to 
G.B. 
\end{acknowledgments}

\end{document}